\documentstyle[11pt,twoside,jltp,psfig]{article}

\title{Trapped Bose--Einstein condensates
at finite temperature: a two--gas model}

\author{R. J. Dodd,\address{Institute for Physical Science and Technology, University of Maryland at College Park,
College Park, MD 20742, USA}
K. Burnett,
\address{Clarendon Laboratory,
Department of Physics,
University of Oxford,
Parks Road,
Oxford OX1 3PU, UK}
Mark Edwards,
\address{Department of Physics,
Georgia Southern University,
Statesboro,
GA 30460, USA} 
and Charles W.\ Clark
\address{Electron and Optical Physics Division, 
National Institute of Standards and Technology, 
Gaithersburg, MD 20899, USA}}

\runninghead{R.\ J.\ Dodd {\em et al.}}{Two--gas description of dilute Bose--Einstein condensates\ldots}

\begin{document}

\begin{abstract}
A simple picture describes
the results of recent treatments of
partially--condensed,
dilute, trapped Bose gases
at temperature
$T > 0$.
The condensate wavefunction is
nearly identical to that
of a $T=0$ condensate with the
same number of
condensate atoms, $N_0$. The cloud of non--condensed atoms
is described
by the statistical mechanics
of an ideal Bose gas in the combined
potentials of the magnetic trap
and the cloud-condensate interaction.
We provide a
physical motivation for
this result,
show how it emerges in the
Hartree-Fock-Bogoliubov-Popov
approximation, and explore
some of its implications
for future experiments.

\end{abstract}

\maketitle


\section{Introduction}

Most recent experiments \cite{JILA,Ketterle,Hulet}
on dilute,
magnetically-trapped, alkali-atom
Bose gases
have viewed phenomena which
are well described
by the zero--temperature ($T = 0$)
mean--field theory of the
Bose--Einstein condensate (BEC),
in which virtually all the gas
in the system resides in the condensed state.
The technique of evaporative cooling \cite{Cool},
used in all such
experiments, grows the BEC
by selective extraction of the non-condensed, ``thermal,''
component of the gas, which is located
at the outer edges of the trap.\cite{CornellNISTJR}
The $T = 0$ mean--field theory has been found to
give a good account
of many BEC properties observed in 
these systems.~\cite{Theorie,Stringari}
New experiments~\cite{finite1,finite2}
have started to probe BECs at
$T > 0$, and so the testing of
alternative finite--temperature
BEC theories has begun.

This paper draws attention to common features
emerging from several independent finite-temperature
theories, which suggest that a relatively simple picture,
which we call the ``two-gas model,'' describes
many of the properties of a dilute Bose gas with
repulsive atomic pair interactions 
(scattering length $a > 0$).
The two gases concerned are the condensate
gas, the intrinsic properties of which
are essentially 
{\em independent} of temperature; 
and the thermal, non-condensed gas, which 
behaves much like an ideal Bose gas 
at temperature $T$ in an effective potential
created by the condensate.  This model emerges
naturally as a limiting case
of the Hartree-Fock-Bogoliubov-Popov
(HFB-Popov)\cite{GriffinHFB,Hutchinson,SlicedBread} and Hartree-Fock (HF)\cite{BERGEMAN}
approximations,
but its features seem also to be manifest in
recent quantum Monte Carlo \cite{Krauth} and 
semiclassical
\cite{Minguzzi} calculations.
The two--gas picture offers some straightforward
implications for interpretation of experiments
and for further development of first--principles theories.

Our evidence for the validity of 
this picture first emerged from 
large--scale numerical
calculations, but its origin can be traced back qualitatively
within the structure of finite--temperature field theory.
In Sec. 2, we show how such a theory
can plausibly lead to a two--gas scenario.  
Section 3 explores some of the implications of the
model.

\section{Two--gas model as a limit of the
HFB-Popov approximation}

The HFB-Popov equations have been derived 
elsewhere\cite{GriffinHFB}
and we merely state the basic equations here.
In the Heisenberg equation of motion, the Bose field
operator, $\hat{\psi}({\bf r})$ is decomposed into a
$c$-number wavefunction $\psi_0({\bf r})$ that describes a
condensate of $N_{0}$ atoms,
and an operator $\tilde{\psi}({\bf r})$ 
describing the non-condensate atoms:
 $\hat{\psi}({\bf r})
= \sqrt{N_{0}}\psi_0({\bf r}) + \tilde{\psi}({\bf r})$. 
In the HFB-Popov approximation, the 
wave function for a condensate of
trapped atoms satisfies a
generalized Gross-Pitaevskii (GP) equation:

\begin{equation}
\left\{ \hat{H_{0}} + 
U_{0} \left[ N_{0}\left|\psi_0({\bf r})\right|^{2} +
2\tilde{n}({\bf r})\right] \right\} \psi_0({\bf r}) 
= \mu \psi_0({\bf r}) ,
\label{gen_gp_eq}
\end{equation}
where 
$\hat{H_{0}} = \frac{-\hbar^{2}}{2 M} \nabla^{2} + 
V_{\rm trap}({\bf r})$ 
is the Hamiltonian for a single
atom of mass $M$ and
position coordinate ${\bf r}$;
the trapping potential
(for cylindrically symmetric systems of current interest)
is given by
$V_{\rm trap}({\bf r}) = M\left(\omega_{\rho}^{2}\rho^{2} + \omega_{z}^{2}z^{2}\right)/2$, with 
$\omega_{\rho}$ and $\omega_{z}$ the
radial and axial angular frequencies of the trap;
$U_{0} = 4\pi\hbar^{2}a/M$ expresses the
binary interaction between atoms; the chemical potential $\mu$, 
interpreted as the work required to add one more atom 
to the {\em condensate}, is treated as an eigenvalue;
and $\psi_0({\bf r})$ is normalized to unity.

The function $\tilde{n}({\bf r})$ is the density of
the non-condensed component of the gas, 
\begin{equation}
\tilde{n}({\bf r}) = 
\sum_{j}\left\{\left[\left|u_{j}({\bf r})\right|^{2}
+ \left|v_{j}({\bf r})\right|^{2}\right]N_{j} +
\left|v_{j}({\bf r})\right|^{2}\right\},
\label{nc_density}
\end{equation}
where
\begin{equation} 
N_{j} = 
\left[\exp\left(\beta E_{j}\right)- 1\right]^{-1}
\label{BEfactor} 
\end{equation}
is the
Bose-Einstein factor, $\beta^{-1} = k_{\rm B}T$ 
and $k_{\rm B}$ is the
Boltzmann constant. The quasi-particle
excitation
energies $E_{j}$ and
amplitudes $u_{j}({\bf r}), v_{j}({\bf r})$
are obtained by
solution of the coupled HFB-Popov equations:
\begin{equation}
\hat {\cal L} u_j({\bf r}) +
N_{0}U_{0}\left|\psi_0({\bf r})\right|^{2}v_j({\bf r}) 
=  E_j u_j({\bf r}) ,
\label{HFB_u}
\end{equation}
\begin{equation}
\hat {\cal L} v_j({\bf r}) +
N_{0}U_{0}\left|\psi_0({\bf r})\right|^{2}u_j({\bf r}) 
= -E_j v_j({\bf r}) ,
\label{HFB_v}
\end{equation}
where $\hat{{\cal L}} \equiv \hat{H_{0}} + 2U_{0}n({\bf r}) - \mu$,
and
$n({\bf r}) = N_{0}\left|\psi_0({\bf r})\right|^{2} + \tilde{n}
({\bf r})$ is the total trapped-atom density.

In simple physical terms, Eqs. (\ref{gen_gp_eq}-\ref{HFB_v})
describe a condensate subject to interaction with
itself and a thermal cloud, with the cloud being
generated by thermal excitations of condensate
quasi--particles (There is also a non-thermal contribution
to this cloud, the so-called ``quantum depletion'' term
represented by the rightmost term of 
Eq. (\ref{nc_density}), but it is much smaller
than the thermal component except near $T=0$).  
To solve these equations for
a given atomic species and trap configuration, we
fix the values of $T$ and $N_0$, and then
determine all other quantities self--consistently,
eventually obtaining the total number of 
trapped atoms, $N$, via 

\begin{equation}
N = \int d{\bf r}~ n({\bf r}) = N_{0} + \sum_{j} N_j
\label{sumoverstates}
\end{equation}
By carrying
out a sequence of such (laborious) calculations, 
we can map out
the $\left\{N,N_{0},T\right\}$ phase diagram of
the interacting Bose gas.  We present elsewhere 
\cite{SlicedBread}
a detailed comparison of the
results of this approach with experimental data
for the $^{87}$Rb condensate at JILA \cite{finite2}; 
for the
temperature--dependent 
quasiparticle excitation energies, HFB-Popov
agrees with experiment to within 5\% for
temperatures from zero up to 65\% of the 
temperature $T_0$ of the phase transition 
for the corresponding ideal gas
(corresponding to thermal fractions from
less than 1\% to about 50\%).  Although at
present there
are considerable discrepancies as
$T \rightarrow T_0$, 
it seems that
HFB-Popov is a useful working theory 
over a significant temperature range. 

Several calculations \cite{Hutchinson,SlicedBread} have
shown that, for current experiments,
the quantum depletion of a small condensate is
negligible, {\em i.\ e.},  
$\int d{\bf r}~\tilde{n}({\bf r})|_{T = 0} = \sum_j  \int d{\bf r} \left|v_{j}({\bf r})\right|^{2} << N_{0}$. 
This justifies use of the approximation
\begin{equation} 
v_{j}({\bf r}) \equiv 0,
\label{v_is_0}
\end{equation} 
which is equivalent to 
the Hartree-Fock approximation used 
by other authors~\cite{BERGEMAN,STOOF}.
If we apply this approximation to
Eq.\ (\ref{HFB_u}) (and neglect the
contribution of $\tilde{n}({\bf r})$ to
$n({\bf r})$), we obtain an ordinary
Schr\"odinger equation for $u_{j}({\bf r})$,
\begin{equation}
\left[ - \frac{\hbar^2}{2 M}  \nabla^2 + 
V_{\rm eff}({\bf r}) \right] u_j({\bf r}) = E_j u_j({\bf r}),
\label{uLSE}
\end{equation}
where the effective potential, given by
\begin{equation}
V_{\rm eff}({\bf r}) = 
V_{\rm trap}({\bf r}) + 
2 N_0 U_0 |\psi_0({\bf r})|^2,
\label{cloudcond}
\end{equation}
is that of the trap modified by the repulsive pair interaction 
between the
thermally--excited atoms and the condensate density.

If we consider
the case of a relatively small thermal fraction, then
we expect to find the condensate localized near the
center of the trap, so that Eq.\ (\ref{cloudcond})
presents the thermal cloud with a trap and repulsive core.  Thus,
at least the low--energy
quasi--particle amplitudes $u_j({\bf r})$
will be expelled from the core, {\it i.\ e.}, they
will have
little overlap with the condensate wavefunction.
This then gives consistency of Eq.\ (\ref{HFB_v})
with our initial approximation, Eq.\ (\ref{v_is_0}).
It also justifies the approximation
that completes our portrayal of a two--gas 
system: because
of the expulsion of quasi--particle amplitudes from
the condensate, we assume that $\tilde{n}({\bf r})$
can be neglected in Eq.(\ref{gen_gp_eq}), so that
the condensate wavefunction is determined by
solving
\begin{equation}
\left[ - \frac{\hbar^2}{2 M}  \nabla^2 +
V_{\rm trap}({\bf r}) + N_0 U_0 |\psi_0({\bf r})|^2 \right] 
\psi_0({\bf r}) = \mu \psi_0({\bf r}),
\label{GP}
\end{equation}
which is just the usual GP equation for a
condensate of $N_0$ atoms at $T=0$.

Thus, these arguments have led us to a simple picture 
in which the finite--temperature Bose system appears 
to be composed of two distinct gases.  One of these
gases, the BEC, is always effectively at
zero temperature, and is described by an
equation which depends only on its own
atomic population, $N_0$, and the trap parameters.  
The other gas, the thermal cloud,
behaves as a normal Bose gas at finite temperature,
sensing the presence of the condensate
through an elastic interaction; it does not
undergo Bose-Einstein condensation itself, but
serves as an atomic reservoir for the BEC.
This resembles the phenomenon of BEC of an ideal gas
in an external potential, except that we account for
interactions in the identification of the ground
state, and in the modification by the condensate
of the external potential exposed to the thermal
cloud. 

\section{Implications of the two-gas model}

The two-gas model provides us with a
straightforward way of computing the phase
diagram of the dilute Bose gas for $T < T_0$.
If there are $N_0$ atoms in the BEC, we
solve Eq. (\ref{GP}) to obtain what we will
call an equivalent zero--temperature condensate (EZC),
{\em i.\ e.} the corresponding $T=0$ condensate 
that contains $N_0$ atoms.  
The EZC solution provides us an orbital 
$\psi_0({\bf r};N_0)$ and chemical potential $\mu(N_0)$;
with these in hand, we can construct 
$V_{\rm eff}({\bf r})$
and find the spectrum of Eq.(\ref{uLSE}).
This procedure, which is independent of 
$T$, gives us the information we need to
compute the equilibrium value of $N$ for 
given values  of $N_0$ and $T$: we evaluate
Eq.~(\ref{sumoverstates}) from Eq.~(\ref{BEfactor}).  

\begin{figure}
\framebox[5in]{\rule[1.125in]{0in}{1.125in}
\psfig{file=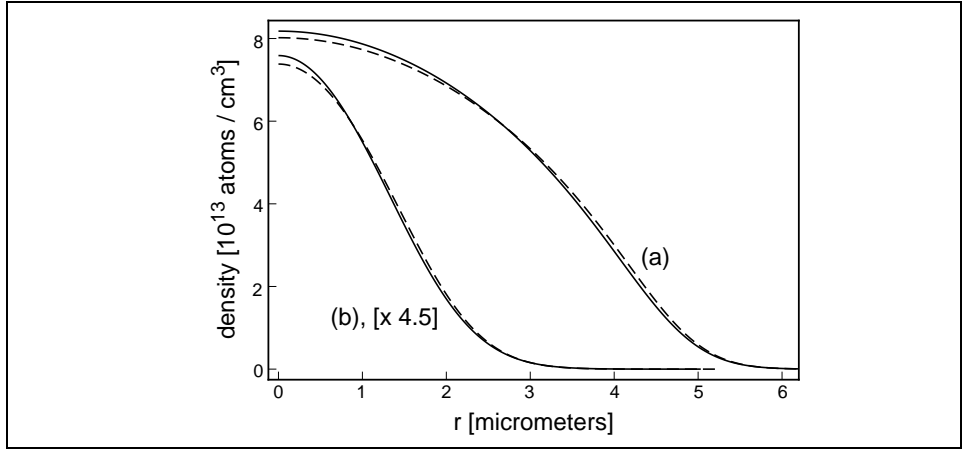,height=2.25in}}
\caption{Condensate densities 
in the $z=0$ plane
for systems of 
$^{87}$Rb atoms in the  
JILA TOP trap~\protect{\cite{JILA}} with radial frequency 
$\nu_\rho = 74$ Hz.  The solid
line shows the condensate density as computed in
the HFB-Popov approximation, and the
dashed line is the density of the
corresponding EZC.
Case (a):  A system of $N = 13150$ atoms
at $T = 70$ nK, corresponding to $f = 0.54$ 
in the HFB-Popov
approximation, 
i.\ e.\  $N_0 = 7106$.
(b) A system of $N= 2000$ atoms at
$T = 51$ nK, corresponding to $f = 0.1$.
The EZC densities are seen to be
very close to those
of the HFB-Popov approximation.}
\label{HFBEZC}
\end{figure}

In short, the EZC provides a ``reference condensate''
which, for a given set of
trap parameters, 
describes all systems with the 
{\em same number 
of condensate atoms} $N_0$.  As we have 
shown elsewhere,~\cite{SlicedBread} this
model provides good agreement with the results of 
full HFB-Popov calculations of condensate and thermal
densities and the critical temperature $T_0$; the
emergence of an EZC can also be seen in the analysis by
Krauth \cite{Krauth} of the results of his 
quantum Monte Carlo calculations, and in
the recent quasi--classical calculations
of Minguzzi {\em et al.} \cite{Minguzzi}.
Comparison of the EZC condensate densities
with those of HFB-Popov calculations is made in 
Fig. (\ref{HFBEZC}); this shows that even in
cases where the condensate fraction
$f =N_0/N$ is as small as 0.1,
the condensate is relatively
unperturbed by the presence of the
thermal cloud.

Another straightforward implication of the
two--gas model concerns the density profile of
the thermal cloud.  If we entertain the simple
hypothesis that the cloud would be described by
classical statistical mechanics of an ideal
gas, then its density $\tilde{n}({\bf r})$
would be proportional to 
$\exp\left[-\beta V_{\rm eff}({\bf r})\right]$.
Since $V_{\rm eff}({\bf r})$ is repulsive at
small $|{\bf r}|$ and confining at large $|{\bf r}|$,
$\tilde{n}({\bf r})$ will attain its maximum
away from the center of the condensate.  If 
we consider the Thomas--Fermi 
limit~\cite{BP}
appropriate to large condensates, then for
the case of a spherical condensate
[$V_{\rm trap}({\bf r}) = M \omega^2 r^2/2$] we find
that 
\begin{eqnarray}
V_{\rm eff}({\bf r}) = \cases{\frac{M \omega^2}{2} \left( 2r_{\rm TF}^2 - r^2 \right), & $r < r_{\rm TF}$ \nonumber \cr
\frac{M \omega^2}{2} r^2,& $r > r_{\rm TF}$
}
\label{TF}
\end{eqnarray}
where the Thomas--Fermi radius, 
$r_{\rm TF}=\left[\frac{15 N_0 U_0}{4 \pi M \omega^2}\right]^{1/5}$,
defines the sharp boundary of the condensate.  
Thus in this limit, $\tilde{n}({\bf r})$ is largest
at the surface of the condensate,
and its distribution becomes more localized
as $N_0$ increases, albeit slowly.
  
\begin{figure}
\framebox[5in]{\rule[1.125in]{0in}{1.125in}
\psfig{file=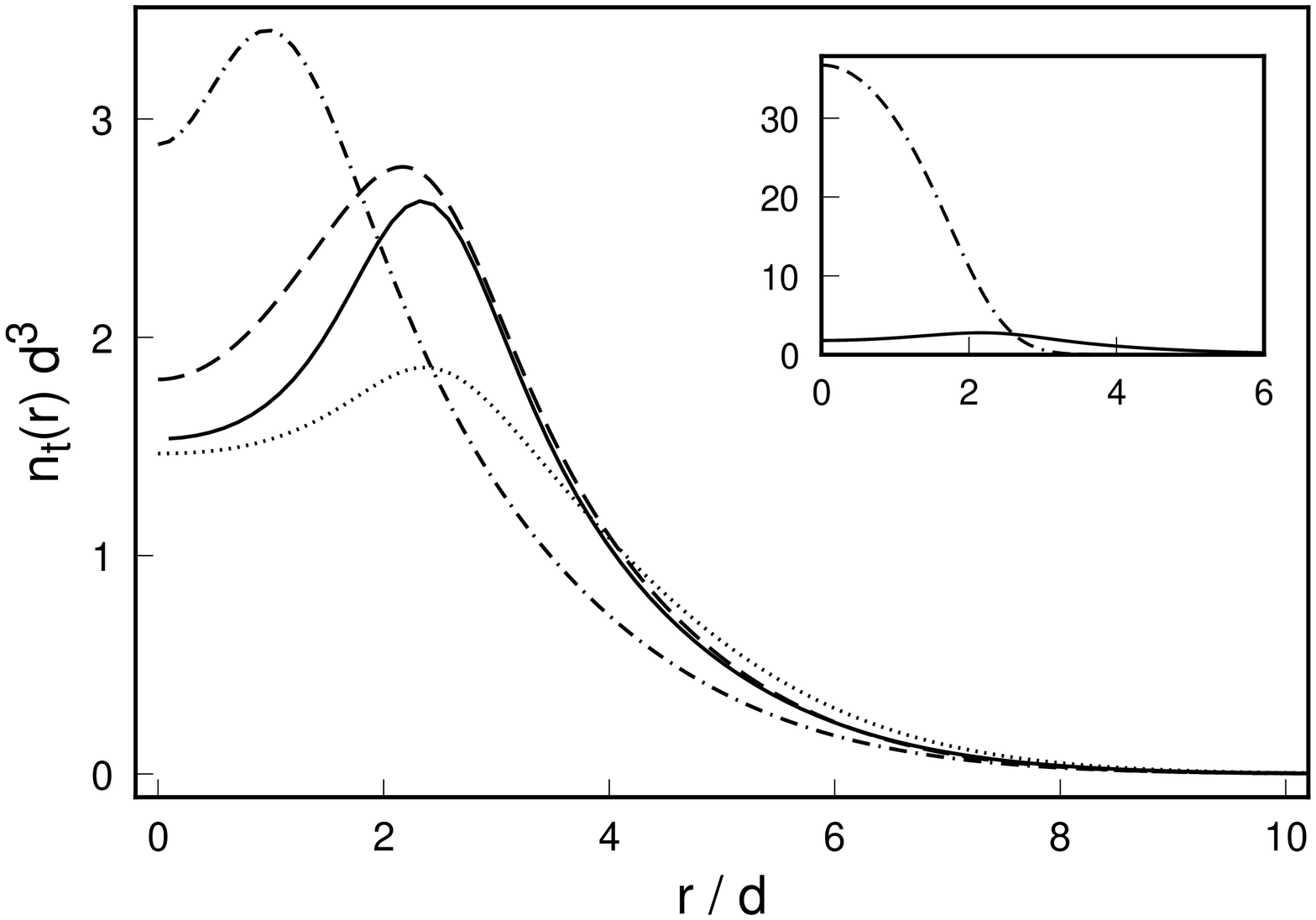,height=2.25in}}
\makebox[5in]{\rule[1.125in]{0in}{1.125in}}
\caption{Thermal density for $N \sim 2000$ 
$^{87}$Rb atoms  
in a spherical trap with 
$\nu_r = \omega/(2\pi) = 200$ Hz, $T=75$ nK, 
corresponding to $f = 0.5$ in the HFB-Popov
approximation. The radius $r$ is given in units
of the characteristic length of the oscillator
$d = \protect{\sqrt{\hbar/(M \omega)}}$.  These parameters
were previously
used by Hutchinson et al.,~\protect{\cite{Hutchinson}} 
whose numerical HFB-Popov results we have
reproduced and use here.  The main figure shows
the thermal density
as computed in 
various approximations.  Chain--dashed line: the
confined ideal quantum gas; dashed line: 
full solution of the HFB-Popov equations; 
solid line: full quantum--statistical distribution
as computed from two--gas model, which obtains
$N=1965$ (vs. the exact value of 2000)
from $N_0 = 1000$ and $T=75$ nK;
dotted line: result of classical statistical
mechanics applied to the two--gas model, 
using a fit to force $N=2000$.  The
inset compares the HFB-Popov thermal density
(solid line) with that of the condensate
(dashed). Thus, even when the system
is only 50\% condensate, the peak condensate
density is clearly much higher than that 
of the thermal cloud.}
\label{THERMAL}
\end{figure}

The key qualitative aspects of this classical
description are applicable to the
quantum system, as shown in Fig.(\ref{THERMAL}):
this displays results of a full 
quantum-mechanical 
finite-number description, without 
any of the semiclassical continuous 
spectrum approximations made 
by other authors \cite{Minguzzi,GIORGINI}.
This figure clearly suggests that 
quantitative interpretations
of experimental data on finite--temperature
condensates, {\em e.g.}
determination of a condensate
fraction from density measurements,
will have to invoke some detailed
model of the thermal distribution, since
this distribution is neither monotonic nor close to
the results obtained for a noninteracting
gas.  On the other hand, our current model
suggests that the condensate and thermal
densities are relatively distinctly segregated
within the cloud, which may considerably simplify
the qualitative understanding of some
properties.  
Since it originates in the distinction
between interactions of condensate and non--condensate
atoms, this spatial segregation of the
two gases seems to be a fundamental aspect
of the behavior of {\em inhomogeneous} Bose gases,
such as the trapped--atom systems of current
interest.  This may have interesting consequences
for applications: for example, it may be possible
to selectively extract condensate {\em vs.} thermal
atoms from a trap by appropriate positioning
of a probe, thus obtaining an outcoupled 
matter wave with higher coherence
than would be otherwise expected.~\cite{G2} 
In homogeneous systems, on the other
hand, condensate and non--condensate populations
are intertwined; this is one of the essential
features of the two--fluid model of liquid
helium.\cite{BEL}

\begin{figure}
\framebox[5in]{\rule[1.125in]{0in}{1.125in}
\psfig{file=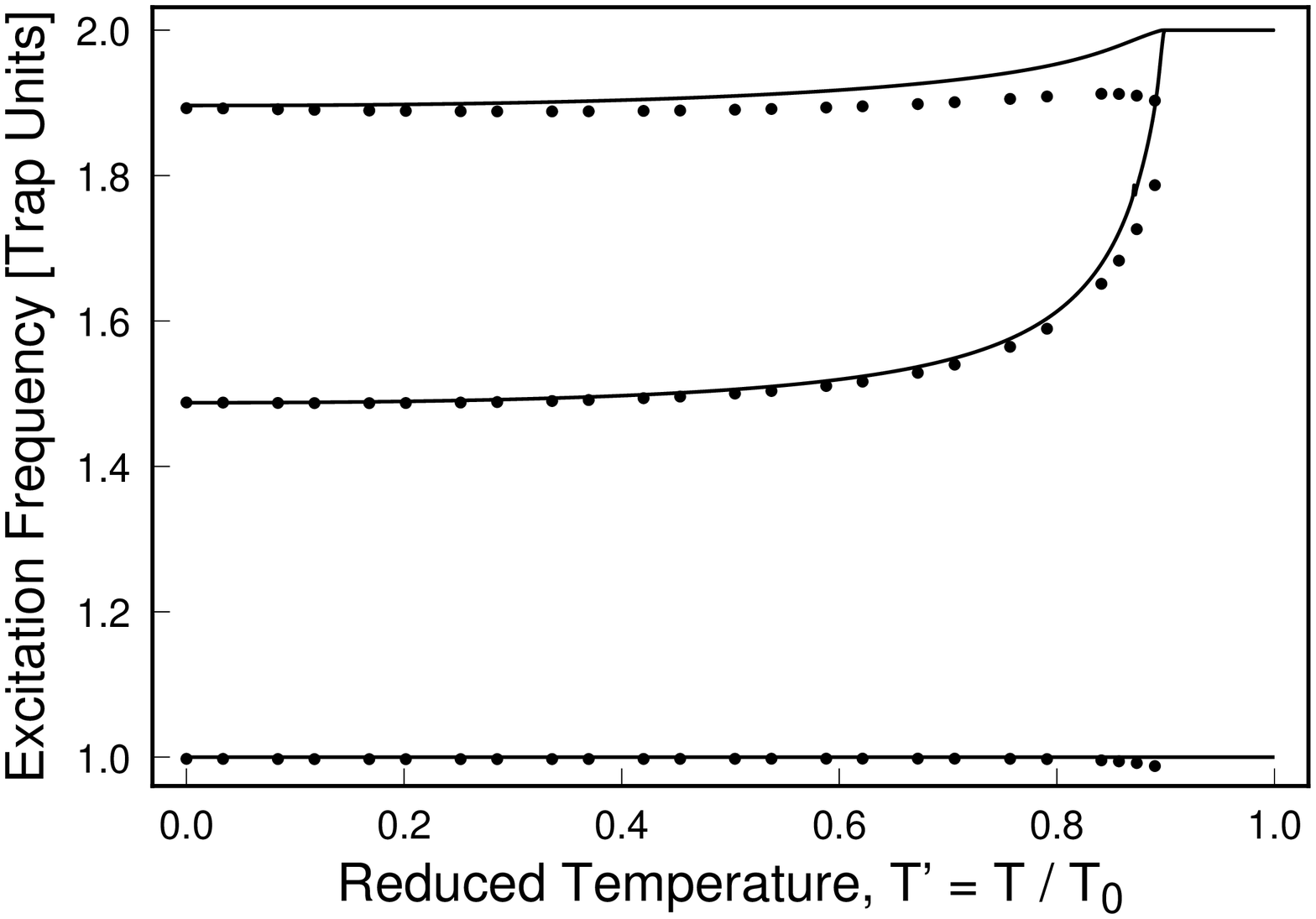,height=2.25in}}
\caption{HFB-Popov excitation frequencies (filled circles)
for the $m=0$ (top), $m=2$ (middle), and the $m=1$ modes
(bottom) for the JILA TOP trap 
with radial frequency $\nu_r =129$~Hz and $N = 2000$ 
$^{87}$Rb atoms, 
{\em vs.} temperature 
in units of $T_0$.
Overlaid (solid lines) are the frequencies
for a zero-temperature system with the same number
$N_0$ of condensate atoms as in the finite-temperature
system.}
\label{frequencies}
\end{figure}

Taking this idea further, we suggest that
any property of a finite--temperature BEC
should be compared in the first instance 
to that of the corresponding EZC.  In the
two--gas model, we expect most of the
$T$-dependence of a given quantity
to be reduced to $N_0$-dependence.  For
example, in Fig. (\ref{frequencies})
we show the quasiparticle excitation 
frequencies for
the JILA TOP trap, over a range of
temperatures relevant to recent experiments,
as computed in the
full HFB-Popov approximation
and in the two--gas model.  It is seen
that the two methods agree up to 
temperatures quite close to the phase
transition, so the main effect of
finite temperature is renormalization
of the value of $N_0$.  An analogous result 
was seen in earlier calculations\cite{SK,PG} for the
{\em homogeneous} Bose--condensed gas
of the temperature dependence of the
speed of sound,
which found it to be given
by an equivalent $T=0$ expression adjusted
for the temperature--dependence of the
condensate density.

\section{Conclusions}

We find that condensate and thermal populations of
a partially Bose--Einstein--condensed trapped--atom
system separate out to a considerable extent.  Treating
the condensate as uncoupled from the thermal cloud, and
the thermal cloud as interacting with a static condensate
potential, yields results similar to those that come
from involved self--consistent field calculations.
These results motivate the identification of the
equivalent zero--temperature condensate (EZC) as a 
consolidating feature of finite--temperature systems.
In this model, the main effect of finite temperature
on the condensate is {\it depletion} of the condensate number. 
Condensate properties that depend only weakly upon
$N_0$, such as the quasi--particle spectrum in the
large--$N_0$ limit
(corresponding to the excitation 
frequencies of large condensates),~\cite{Stringari}
should exhibit only weak
temperature dependence.

\section*{ACKNOWLEDGMENTS}
This work was supported in part
by the U.\ S.\ National Science Foundation under grants PHY-9601261 and
PHY-9612728, the U.\ S.\ Office of Naval Research, 
by MCS grant No. PAN/NIST-93-156,
and by the
U.\ K.\ Engineering and Physical Sciences Research Council.


\end{document}